\documentclass[aip, amsmath,amssymb, reprint, author-numerical,]{revtex4-1}

\usepackage{dcolumn}
\usepackage{bm}

\usepackage[T1]{fontenc}
\usepackage{mathptmx}
\usepackage{bbold}
\usepackage[utf8]{inputenc}
\usepackage{natbib}
\usepackage{graphicx}
\usepackage{floatflt}
\usepackage[hidelinks]{hyperref}
\usepackage[normalem]{ulem}
\usepackage{url}

\usepackage{xcolor}

\newcommand{\rev}[1]{{\color{black}{#1}}}

\usepackage{mathtools}

\newcommand{\fig}{Fig.}

\makeatletter
\let\@fnsymbol\@fnsymbol@latex
\@booleanfalse\altaffilletter@sw
\makeatother

\draft 

\begin{document}

\title{Obscuring digital route choice information prevents delay-induced congestion}

\author{Verena Krall}
\altaffiliation{These authors contributed equally.}
\affiliation{Chair for Network Dynamics, Center for Advancing Electronics Dresden (cfaed), Technical University of Dresden, 01069 Dresden, Germany}
\affiliation{Institute for Theoretical Physics, Technical University of Dresden, 01069 Dresden, Germany}

\author{Max F. Burg}
\altaffiliation{These authors contributed equally.}
\affiliation{Institute for Theoretical Physics, University of Tübingen, 72076 Tübingen, Germany}
\affiliation{Bernstein Center for Computational Neuroscience, Tübingen, 72076 Tübingen, Germany}

\author{Friedrich Pagenkopf}
\affiliation{Institute for Theoretical Physics, Technical University of Dresden, 01069 Dresden, Germany}

\author{Henrik Wolf}
\affiliation{Institute for Theoretical Physics, Technical University of Dresden, 01069 Dresden, Germany}

\author{Marc Timme}
\affiliation{Chair for Network Dynamics, Center for Advancing Electronics Dresden (cfaed), Technical University of Dresden, 01069 Dresden, Germany}
\affiliation{Institute for Theoretical Physics, Technical University of Dresden, 01069 Dresden, Germany}
\affiliation{Lakeside Labs, 9020 Klagenfurt, Austria}

\author{Malte Schröder}
\email[Electronic Address: ]{malte.schroeder@tu-dresden.de}
\affiliation{Chair for Network Dynamics, Center for Advancing Electronics Dresden (cfaed), Technical University of Dresden, 01069 Dresden, Germany}
\affiliation{Institute for Theoretical Physics, Technical University of Dresden, 01069 Dresden, Germany}

\date{\today}
\begin{abstract} 
Although routing applications increasingly affect individual mobility choices, their impact on collective traffic dynamics remains largely unknown. Smart communication technologies provide accurate traffic data for choosing one route over other alternatives, yet inherent delays undermine the potential usefulness of such information. Here we introduce and analyze a simple model of collective traffic dynamics which result\rev{s} from route choice relying on outdated traffic information. We find for sufficiently small information delays that traffic flows are stable against perturbations. However, delays beyond a bifurcation point induce self-organized flow oscillations of increasing amplitude -- congestion arises. Providing delayed information averaged over sufficiently long periods of time or, more intriguingly, reducing the number of vehicles adhering to the route recommendations may prevent such delay-induced congestion. We reveal the fundamental mechanisms underlying these phenomena in a minimal two-road model and demonstrate their generality in microscopic, agent-based simulations of a road network system. Our findings provide a way to conceptually understand system-wide traffic dynamics caused by broadly used non-instantaneous routing information and suggest how resulting unintended collective traffic states could be avoided.
\end{abstract}

\maketitle 

\begin{quotation}
App-based traffic density and travel time information increasingly influence\rev{s} route choice decisions of drivers. However, individual routing decisions, in turn, change the traffic density. Here we study how this feedback loop affects the collective traffic state. We find that if routing information is delayed, systemic inefficiencies may arise as drivers collectively react to a previous state of the system without taking recent changes into account. Such delayed reactions may induce transient oscillations of the traffic flow between two alternative routes. Once the delay exceeds a critical value\rev{,} the equilibrium flow becomes unstable. The oscillations grow until both routes congest. We present two possibilities to reduce this undesired effect of delayed information. First, instead of providing information about a traffic state at a single time-point, averaging travel time information may reduce the impact of temporary travel time peaks on routing decisions and thus prevent congestion due to collective overreaction. Second, withholding travel time information from some drivers may, intriguingly, counterbalance the delayed (over)reaction of others, thereby also decreasing the risk of congestion.
\end{quotation}

\section{Introduction}

Digitization, real-time collection of data and access to information have transformed everyday life. For example, drivers nowadays often base their route choices on traffic information accessed en-route in almost real-time. In principle, choosing and adapting the travel route based on current travel times might lead to more efficient usage of the street infrastructure and help prevent congestion. However, this notion is controversial as providing information about the travel time is not necessarily beneficial for the overall state of the system and may even increase congestion~\cite{ben-akivaDynamicNetworkModels1991, depalmaRiskAversionValue2012, lindseyPretripInformationRoutechoice2014, rapoportPretripInformationRoutechoice2014, tavafoghiInformationalIncentivesCongestion2017, storch_traffic_2020}, for example, if the available information is imperfect~\cite{arnottDoesProvidingInformation1991}. Specifically, it has been shown from a systemic perspective that selfish routing may induce non-optimal collective states in which the travel time averaged across all vehicles is higher than the theoretical optimum~\cite{braess_uber_1968, braessParadoxTrafficPlanning2005, colak_understanding_2016, carrasco_does_2020, storch_traffic_2020}. Furthermore, unpleasant side-effects exist, such as increased usage of low-capacity roads through residential areas, use of complicated routes with higher accident risk, and increased noise and air pollution~\cite{j_thai_negative_2016, johnsonBeautifulWhatCost2017, festaModelingImpactOnline2019a}. 

In addition to unintended negative effects of wide-spread usage of routing apps, the travel time information that drivers rely on to select their route is inherently outdated~\cite{yokoya_dynamics_2004, scellato_traffic_2010}. The information only reflects temporary traffic states and relevant decisions of other drivers are not taken into account. Once all drivers choose the route with the predicted shortest travel time based on outdated information, it might not be the fastest route and unforeseen congested states may arise by the time they are en-route. This kind of collective behavior is termed as \emph{overreaction}~\cite{ben-akivaDynamicNetworkModels1991} and may induce oscillations in route utilization~\cite{horowitz_stability_1984, wahle_decision_2000, wang_advanced_2005, nie_equilibrium_2010, imai_optimal_2015}.

Here, we analyze a simple macroscopic traffic flow model, describing how individual route choice decisions based on delayed information may impact the collective traffic dynamics. We investigate the emergence of traffic flow instabilities due to delayed information and how to prevent the resulting congestion. Studying the mechanism underlying such instabilities for a two-road system reveals a \rev{Hopf-}bifurcation where the stability of the equilibrium flow changes: above a critical delay of travel time information, street load oscillations increase in amplitude, eventually causing a traffic jam throughout the system. Inspired by findings that averaging information prevents destabilizing feedback loops, for example in decentral smart power grids~\cite{schafer_decentral_2015} and information routing in optical fiber networks~\cite{peroza_marval_adaptive_2011}, we demonstrate how congestion induced by delayed traffic information may be prevented by offering time-averaged rather than instantaneous travel time predictions. We confirm the robustness of our analytical insights with microscopic agent-based simulations in a larger network. We furthermore show that the same mechanism of avoiding overreaction prevents congestion if only a fraction of all drivers has access to the information -- a counterintuitive finding given that providing travel time information is supposed to decrease driving times.

\section{Results}

\subsection{Minimal model system}

\begin{figure}[h!]
    \begin{center}
    \includegraphics[width=8.5 cm]{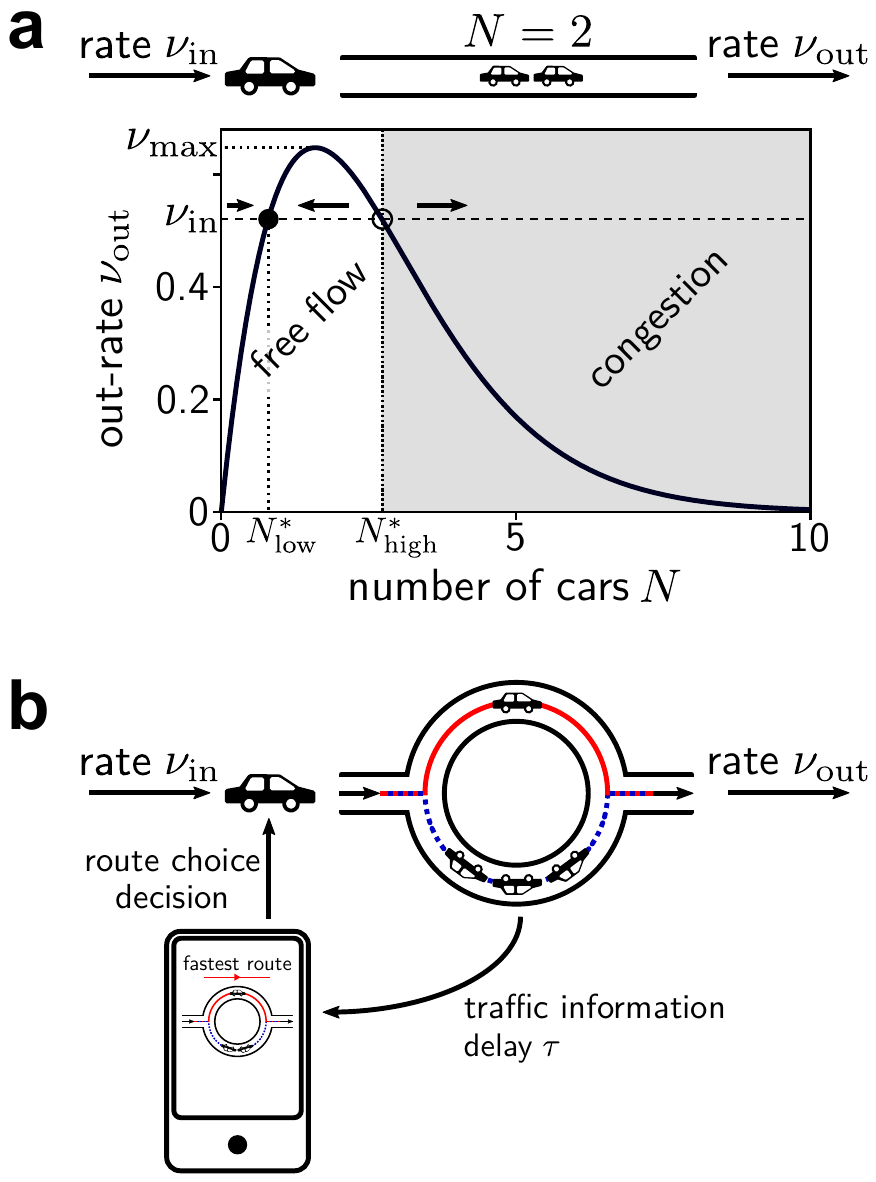}
    \end{center}
    \caption{\textbf{Route-choice model.}
    (a) Vehicles enter a street with a rate $\nu_\text{in}$ and leave it with a rate $\nu_\text{out}$\,. The flow across the street depends on the current traffic conditions on the route [Eq.~\eqref{outrate}]: \rev{the system [Eq.~\eqref{DGL}] settles} either into a stable fixed point $N^*_\mathrm{low}$ or becomes congested once more than $N^*_\mathrm{high}$ vehicles are on the street (congestion, grey shading).
    (b) Among multiple route options, drivers tend to choose the fastest route [Eq.~\eqref{eq:decision}] based on the travel time information provided by, for example, a routing app. 
    However, the \rev{available} travel time estimates reflect the traffic conditions with a delay $\tau$ due to measurement and updating delays of the information or before they become relevant to the driver later during the trip. In this time span, the traffic conditions may change due to unpredictable decisions of other drivers.
    } 
    \label{fig:model}
\end{figure}

The macroscopic state of traffic along a single street at time $t$ is described by the number of vehicles $N(t)$ which changes as vehicles enter \rev{the street at constant rate $\nu_{\text{in}}$ and leave the street at rate $\nu_{\text{out}}(N)$, depending on the current traffic conditions. The time evolution of $N(t)$ thus satisfies the differential equation }
\begin{align}
    \frac{\mathrm{d} N}{\mathrm{d} t} &= \nu_{\text{in}}-\nu_{\text{out}}(N)\ . \label{DGL}
\end{align}

On a street with a capacity $N_0$\, the distance between two vehicles \rev{decreases as the number of vehicles increases}, which forces drivers to slow down. Thus, the average velocity of vehicles on a street decreases, which is equivalent to an increase of the travel time compared to the time $t_0$ a vehicle needs to drive through the empty street,
\begin{align}
    t_\text{travel}(N)=t_{0} \left(\frac{N}{N_0} \right)^{-1} \left(\exp\left(\frac{N}{N_0}\right)-1\right) \quad, \label{traveltime}
\end{align}
i.\,e. the travel time grows exponentially with the number of vehicles. 
Such an exponential dependence has been suggested in early studies of traffic flow \cite{smock1962iterative} and correctly captures the qualitative dependence of travel time. For small numbers $N/N_0 \rightarrow 0$, this function resembles the most common travel time function used \cite{branston_link_1976}\,. 

In this macroscopic description, the instantaneous rate of outflow is given as the number of vehicles currently on the street divided by the current travel time
\begin{align}
    \nu_{\text{out}}(N)&= \frac{N}{t_\text{travel}(N)}\nonumber\\
    &=\frac{N^2}{t_{0} N_{0}}\left\lbrack\exp\left(\frac{N}{N_{0}}\right)-1\right\rbrack^{-1} \quad . \label{outrate}
\end{align}
\rev{This function qualitatively resembles the form of fundamental density-flow-diagrams of traffic flow (\fig~\ref{fig:model}a) found empirically\cite{nagel_cellular_1992}, where the flow across the street depends on the current traffic conditions on the route.} 
\rev{The rate of outgoing vehicles increases approximately linearly for few vehicles on the route, saturates at intermediate numbers of vehicles because the velocity of vehicles decreases, and finally $\nu_\text{out}$ decreases as $N>N(\nu_\text{max})$. 
For each given subcritical rate of incoming vehicles $\nu_\text{in} < \nu_\text{max}$\,, the right-hand side of the differential equation \eqref{DGL} equals zero for two street load values: 
$N=N^*_\mathrm{low}$ is a stable fixed point, such that systems with street loads up to $N^*_\mathrm{high}$ will rapidly relax to this free flow state. 
The unstable fixed point $N^*_\mathrm{high}$ describes the onset of congestion: Once the street load surpasses this value even by a small amount, fewer vehicles leave the street than enter it due to the increased travel time -- the street becomes congested (\fig~\ref{fig:model}a). If the rate of incoming vehicles exceeds the capacity $\nu_\mathrm{max}$, no stable fixed point exists and the street always becomes congested.}

\rev{We remark that our overall findings on the} impact of outdated information \rev{do} not depend on the exact curve given by Eq.~\eqref{outrate} but rather on its qualitative form. Our choice for equation~\eqref{traveltime} thus does not put any limits on the generality of our findings. 

If drivers have various route options $i$ to reach their destination, \rev{they tend to choose the fastest route based on the travel time information $t_\text{signal}$ provided by, for example, a routing app (\fig~\ref{fig:model}b). Hence,} the inflow rate $\nu_{\text{in},i}$ of each of the routes $i$ depends on the drivers' route choices. Given that travel times $t_{\text{signal},i}$ for all streets $i$ are provided to the drivers, it has been found empirically that they choose one route $i$ out of a set of route options indexed by $j$ probabilistically, following a Boltzmann distribution as described by the multinomial logit model \citep{domencich_urban_1975, helbing_dynamic_2004}
\begin{align}
    \nu_{\text{in},i} &= \nu_{\text{in}}^0 \left(\frac{e^{-\beta t_{\text{signal},i}}}{\sum_j e^{-\beta t_{\text{signal},j}}}\right) \ , \label{eq:decision}
\end{align}
where $\nu_{\text{in}}^0 = \sum_i \nu_{\text{in},i}$ is a constant denoting the total rate of vehicles entering the road-system and $\beta$ is a parameter which governs the impact of information on the individual decision\rev{s}: $\beta=0$ corresponds to uniformly random choices, $\beta\rightarrow\infty$ describes deterministic best response dynamics. Within the following, we set an intermediate value $\beta=1$\,, such that drivers react sufficiently to travel time differences but route choices are not fully deterministic. 

\rev{First, to illustrate the basic effect of outdated information, we study a minimal} two-road network, $i \in \{1,2\}$, with identical normalized length and capacities, $N_{0,i}=t_{0,i}=1$\,, and interpret $N_i$ as the density of vehicles on street $i$. \rev{The complete system of (delayed) differential equations governing the street load dynamics is given in Eq.~\eqref{complete_model}\,.}  \rev{To investigate the impact of outdated traffic information on the traffic dynamics, we consider here a small initial deviation from the steady state $N^\ast_\text{low}$ on both streets. We note that the number of vehicles on both streets will rapidly settle at the fixed point value $N^\ast_\text{low}$ even in the absence of information (compare Fig.~\ref{fig:model}a) due to the symmetry of this minimal setting. However, the setting is sufficient to observe the fundamental impact of delay compared to current information on traffic flow stability. We illustrate the robustness of our results in a larger, heterogeneous network in section \ref{network_section} below, where up-to-date traffic information does improve traffic flow.}

\subsubsection{Stable equilibrium flow when providing up-to-date information}

\begin{figure*}[hbt!]
   \begin{center}
       \includegraphics[width=17 cm]{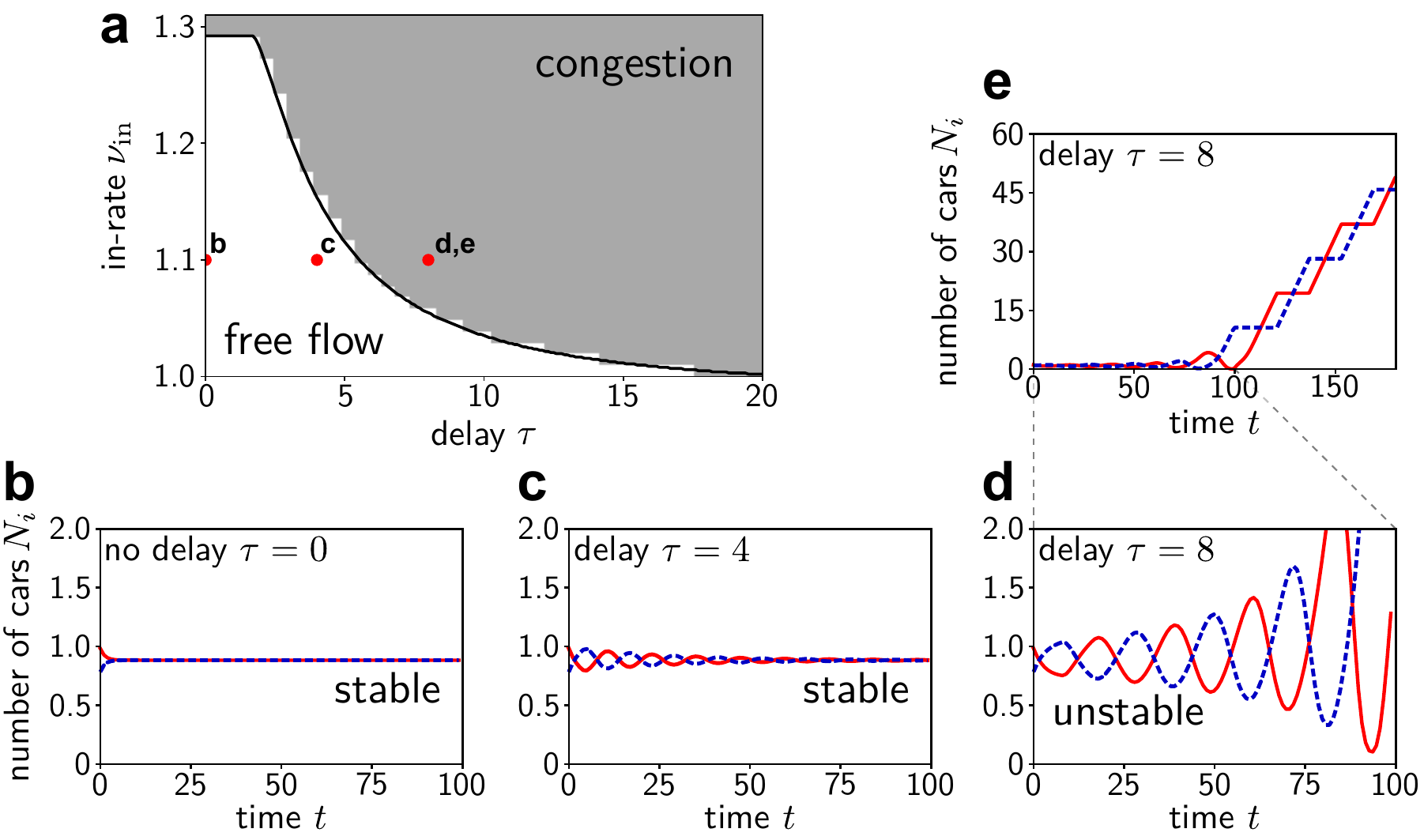}
   \end{center}
   \caption{
   \textbf{Out-dated traffic information induces congestion}. 
  \rev{(a) 
  Simulations show conditions under which a small initial imbalance of the load on both streets induces congestion} (grey shading), confirmed by the linear stability analysis (black line, see appendix \ref{LSA_Appendix}). 
  (b-e) Time evolution of the street load for increasing delay (across panels) at identical in-rate $\nu_\mathrm{in} = 1.1$. The colors of the graphs \rev{indicate} the two different routes in the model shown in \fig~\ref{fig:model}b.
  \rev{For small information delay, the amplitude of the oscillations decrease and the system settles to the stable fixed point (panel~b,~c). If the information delay increases above a critical delay, oscillations grow until both streets become congested (panel~d,~e). \rev{All results are based on Eq.~\ref{complete_model} with parameters $t_{0,i}=N_{0,i}=1$\,, $\beta=1$ and initial street loads $N_i(t\leq 0) = N^*_\text{low}(\nu_{\text{in}}) \pm 0.1$\,, with $N^*_\text{low}(\nu_{\text{in}}=1.1)\approx 0.884$ and $N^*_\text{high}(\nu_{\text{in}}=1.1)\approx 2.554$\,.}}
   \label{fig:stabdelay}
   }
\end{figure*}

If drivers are provided with up-to-date information on the travel times for both routes, their route choice is a reaction to current imbalances in street loads. Hence, the two-road system will only become congested if the inflow rate of vehicles is higher than the outflow rate. The maximal in-rate of a single street that does not lead to congestion, $\nu_\text{max}$\,, is given by the maximum of the out-rate $\nu_\text{max} = \max_N \nu_\text{out}(N)$ (\fig~\ref{fig:model}a). \rev{For the minimal} two road system, traffic will flow freely for in-rates smaller than $2\nu_\text{max}$ if no information delay exists (\fig~\ref{fig:stabdelay}a). The density of vehicles on both route options will settle at a stable equilibrium \rev{(\fig~\ref{fig:stabdelay}b)}, which is given by the fixed point $N^\ast_\text{low}$ marked in \fig~\ref{fig:model}a. Once the inflow exceeds the critical in-rate of the two-street network, the traffic dynamics undergo a \rev{saddle-node bifurcation} and become congested.

\subsubsection{Unstable traffic flow due to outdated travel time information}

\rev{In a more realistic scenario, the available travel time estimates reflect traffic conditions with a delay $\tau$ due to measurement and updating delays of the information or because they become relevant to the driver later during the trip. During this time span, the traffic conditions may change due to unpredictable decisions of other drivers.}
\rev{The delay-differential equation \cite{driver_ordinary_1977} that governs the dynamics of street loads is then
\begin{align}
    \frac{\mathrm{d} N_i}{\mathrm{d} t} &= \nu_{\text{in},i}\lbrack N_1 (t-\tau), N_2(t-\tau)\rbrack -\nu_{\text{out},i} \lbrack N_1(t), N_2(t)\rbrack \ . \label{DDE}
\end{align}}

The general effect of route choice based on delayed information is a change in stability of the two-road network dynamics, as \rev{characterized by the bifurcation diagram \fig~\ref{fig:stabdelay}a which illustrates under which conditions of in-rate and time delay a small initial imbalance of the load on both streets leads to congestion.} While for small delays the maximal in-rate is still only bounded by the street capacities, with increasing delay congestion sets in already for in-rates which are below the maximal value. 

If drivers receive information about travel times based on measurements at an earlier time $t-\tau$\,, drivers keep choosing a route which was optimal at the moment of measurement but might not be the fastest route anymore. As a result, the road that seemed better at measurement time becomes even more crowded than the formerly worse alternative\rev{, leading to oscillations of the street load (\fig~\ref{fig:stabdelay}c-e).} Only when this change in relative street loads becomes visible in the delayed information, drivers adapt their route choice behavior.

As long as the information delay $\tau$ is smaller than a critical delay $\tau_c(\nu_\text{in}^0)$ the load differences between the two roads stay small such that some drivers keep choosing the seemingly worse option. Moreover, the rate of outgoing vehicles is sufficiently high even for maximal loads to counterbalance the extreme values of street load, as shown in \fig~\ref{fig:stabdelay}c\,: the number of vehicles on each street oscillates around the stable fixed point $N^\ast_\text{low}$ with decreasing amplitude\rev{, eventually settling to the symmetric free flow state.}

\rev{However, if the information delay $\tau$ increases above the critical value, the fixed point $N_\text{low}^\ast$ undergoes a Hopf-bifurcation and becomes unstable.} Drivers choose one route preferentially for longer before the travel time information updates. The larger travel time differences cause them to overreact even stronger in the other direction, reinforcing the dynamics and leading to oscillations of the street load with growing amplitude, as illustrated in \fig~\ref{fig:stabdelay}d. Finally, once the number of vehicles on one of the two streets exceeds the second, unstable fixed point $N_\text{high}^\ast$\,, the outflow of vehicles exceeds the inflow and congestion arises. As a consequence, \rev{almost} every incoming driver now chooses the alternative road, which will thus eventually also congest. \rev{In summary, the street load oscillations (\fig~\ref{fig:stabdelay}d) grow until both streets become congested (\fig~\ref{fig:stabdelay}e) -- a long-term effect of large information delays.} 

The critical delay $\tau_c(\nu_\text{in}^0)$ for a given in-rate derived from \rev{linear stability analysis (\fig~\ref{fig:stabdelay}a)} of Eq.~\eqref{DDE} agrees well with the stability regimes obtained from the direct numerical integration of the delay-differential equation~\eqref{DDE} (see \fig~\ref{fig:stabdelay}a\,; for details see Appendix \ref{LSA_Appendix}).

\subsubsection{Averaging over outdated travel-time information prevents congestion}

\begin{figure}[h!]
    \begin{center}
    \includegraphics[width=8.5 cm]{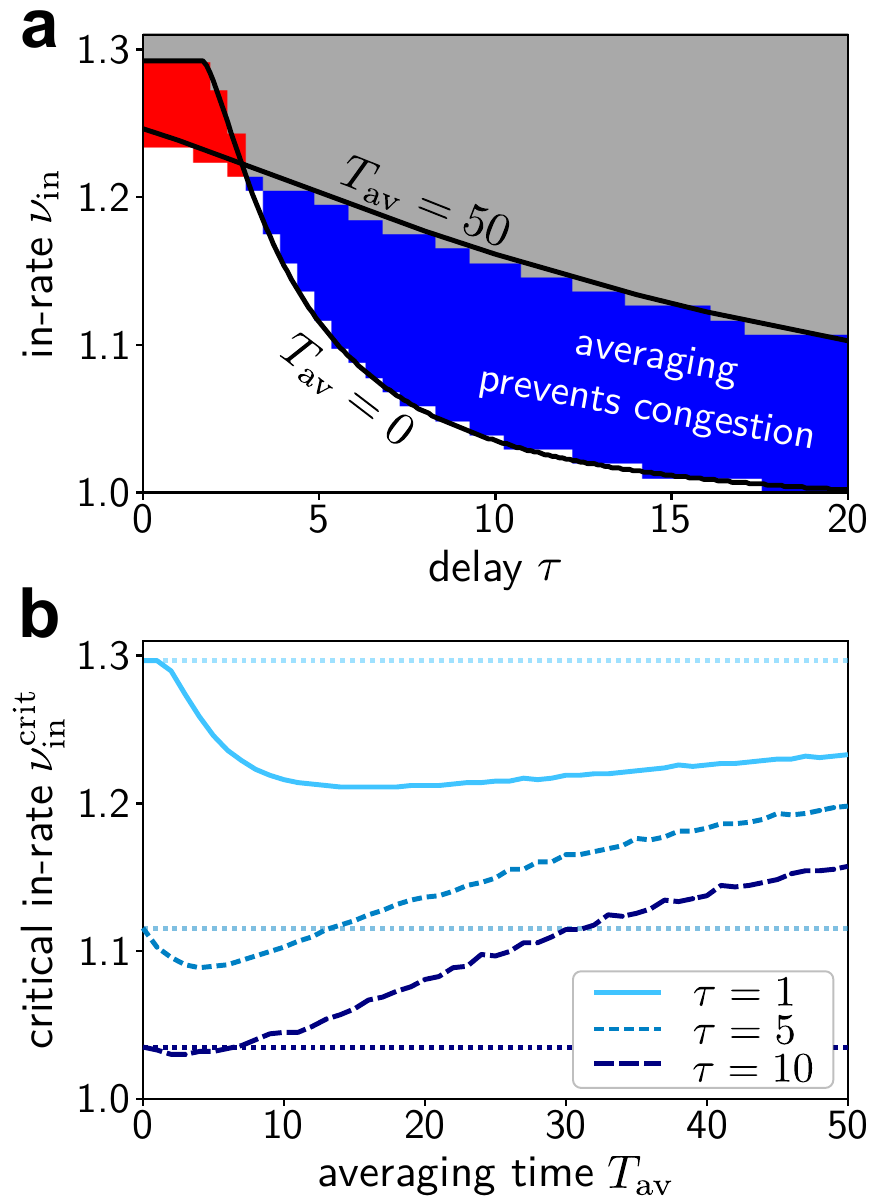}
    \end{center}
    \caption{
    \textbf{Time-averaged information \rev{stabilizes the free flow state at high delays.}}
    (a) \rev{Conditions under which averaged information ($T_\text{av} = 50$) prevents congestion (blue) or even destabilizes the free flow state (red), compared to no averaging (cf. \fig~\ref{fig:stabdelay}a).} The black lines indicate critical in-rates \rev{obtained} from linear stability analysis of the delayed differential equations Eq.~\eqref{DDE} (see Appendix \ref{LSA_Appendix} for further details).
    (b) The effective capacity $\nu_\text{in}^\text{crit}$ before congestion emerges increases with averaging time when the route choice of drivers is based on outdated information. \rev{Dotted lines: no averaging.}
    \rev{While for low delays $\tau$ or low averaging times $T_\text{av}$ there is no beneficial effect, at high delays ($\tau = 10$), averaging induces an increase in the critical in-rate already for small $T_\text{av}$.}
    }
    \label{fig:stabchange}
\end{figure}

Outdated information induces congestion due to an overreaction of drivers to the travel time information they \rev{receive}. To prevent this fatal overreaction, we reduce the differences in the signaled travel times of the two route options by providing travel time information based on the average load,
\begin{align}
    t_{\text{signal},i}=t_{\text{travel}}\left(\langle N_i(t-\tau)\rangle_{T_\text{av}}\right) \ , \label{Averaging}
\end{align}
where 
\begin{equation}
    \langle N_i(t-\tau)\rangle_{T_\text{av}} =\frac{1}{T_\text{av}}\int_{t-\tau-T_\text{av}}^{t-\tau} N_i(t') \, \mathrm{d} t'
    \label{eq:av_load}
\end{equation} 
and $T_\text{av}$ is the length of the time span within which data for the averaged information is collected. \rev{The complete system of differential equations is provided in the appendix in Eq.~\eqref{complete_model_averaging}.}
 
Providing information averaged over past data to drivers, the traffic flow dynamics change as shown in the bifurcation diagram \ref{fig:stabchange}a. For a small delay, i.e. for almost up-to-date information, congestion arises for smaller in-rates compared to no averaging, marked by the red region in \fig~\ref{fig:stabchange}a. This points to the fact that once we average, we effectively introduce additional delays which cause overreactions instead of preventing them. However, in the region of intermediate and high delay, averaging reinstates stability for some in-rates (see the blue region in \fig~\ref{fig:stabchange}a). By averaging over data from a larger time span, extreme values of street loads impact route choice less compared to providing information from just one time point. \rev{Our simulation results are in good agreement with the predictions from linear stability analysis of the corresponding delayed differential equations (\fig~\ref{fig:stabchange}a black lines; see Appendix for further details).}

This stabilizing effect of averaged information depends on the averaging time window $T_\text{av}$ (\fig~\ref{fig:stabchange}b).
\rev{For (almost) up-to-date information ($\tau = 1$), averaging always decreases the capacity of the system compared to no averaging, increasing instability. As the traffic information is more and more delayed, providing time-averaged information counteracts the destabilizing effect. At intermediate delays ($\tau = 5$), only sufficiently large averaging times increase the capacity of the system preventing congestion, while at high delays ($\tau = 10$), averaging almost never has negative effects.}

\subsection{Traffic on networks \label{network_section}}

\begin{figure*}[hbt!]
    \centering
    \includegraphics[width=17 cm]{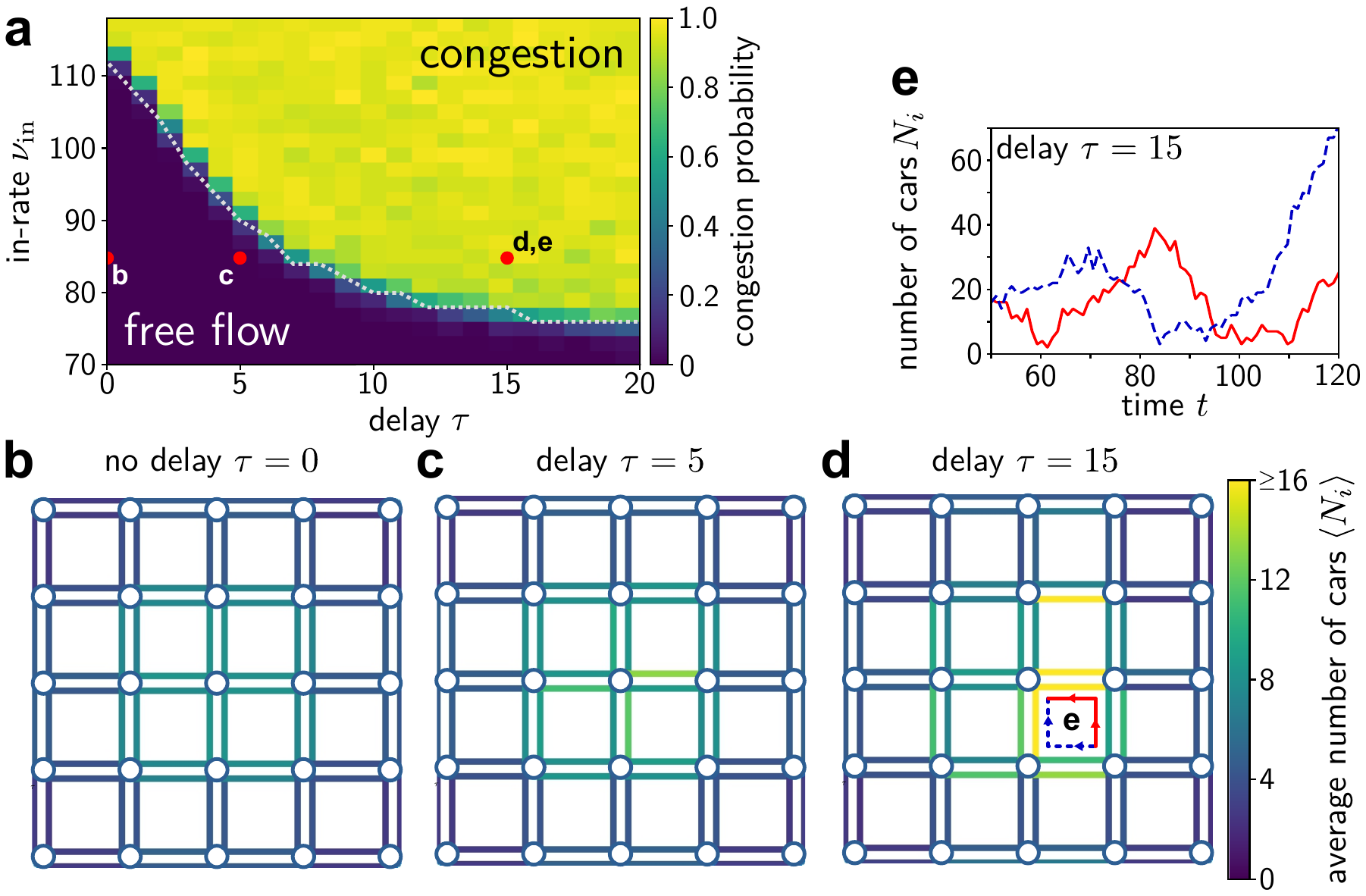}
    \caption{
    \textbf{Out-dated traffic information induces congestion in larger street networks\rev{, extending the results of the two-street model.}}
    (a) Fraction of $100$ \rev{stochastic, agent-based} simulations that became congested for an initially empty square lattice \rev{(compare \fig~\ref{fig:stabdelay}a; see main text for details).} \rev{Dashed line:} boundary where $50\%$ of the simulations end in a congested state.
    (b-d) Time-averaged load of all \rev{undirected} streets in the network, represented by two parallel, directed edges. The central streets in the network are \rev{naturally} utilized more than streets on the boundary. For increasing delay $\tau$, \rev{street load increases up to overload when the delay becomes too large.}
    (e) The mechanism leading to congestion is the same as for the two-street model
    \rev{, showing oscillatory dynamics with growing amplitude until both paths highlighted by red and blue paths in panel~(d) overload.}} 
    \label{fig:fig2_gridversion}
\end{figure*}

\begin{figure}[h!]
    \centering
    \includegraphics[width=8.5 cm]{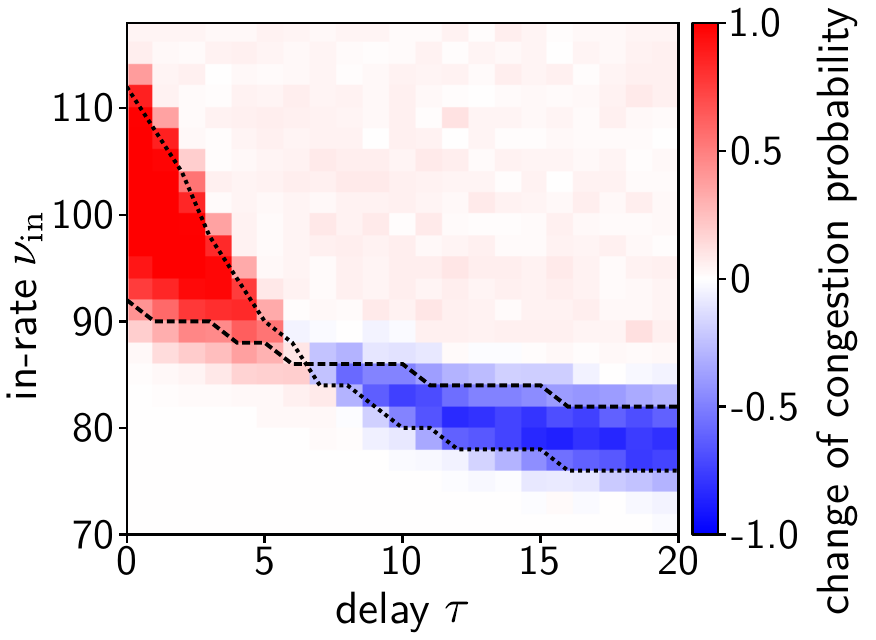}
    \caption{
    \textbf{Time-averaged information prevents congestion at large delays in networks.}
    As in the \rev{minimal} two-street model, time-averaging of delayed traffic information ($T_\text{av} = 50$) stabilizes the free flow state at high delays \rev{(blue, a smaller fraction of realizations resulting in congestion with time-averaging compared to no averaging, cf. \fig~\ref{fig:fig2_gridversion}a)}.
    For small delays, averaging destabilizes the free flow state (red, a larger fraction of realizations result in congestion). \rev{The dashed and dotted lines indicate parameters for which $50\%$ of simulations lead to congestion.}
    }
    \label{fig:fig3_gridversion}
\end{figure}

\begin{figure}[h!]
    \centering
    \includegraphics[width=8.5 cm]{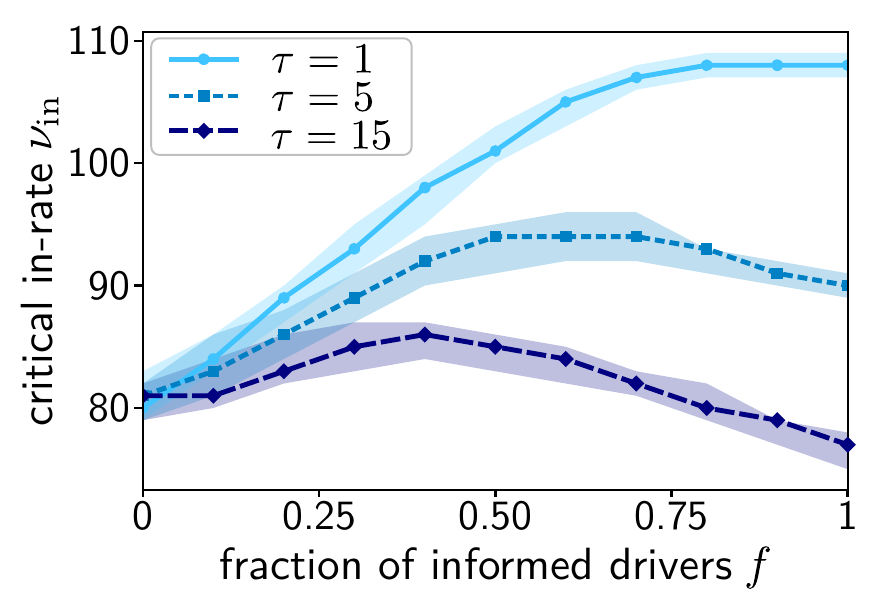}
    \caption{
    \textbf{Intermediate fractions of informed drivers optimize capacity.}
    If only a fraction $f$ of drivers choose routes based on traffic information 
    \rev{ and the remaining drivers choose routes uniformly randomly, the critical in-rate increases for some values of $f$ for intermediate and high delay ($\tau \in \{5, 15\}$), but not for almost up-to-date information ($\tau=1$). Data-points indicate critical in-rates where 50 \% of simulation runs become congested. The shaded regions indicate intervals for which $25\%$ to $75\%$ of runs become congested.}
    } 
    \label{fig:fig4_gridversion}
\end{figure}

To demonstrate the robustness of the analysis from the macroscopic traffic model, i.e. the minimal two-road system, we analyze the impact of delayed information for larger, heterogeneous networks with microscopic agent-based simulations of individual vehicles in a 5x5-square grid. Within the simulation, we set the capacity measure $N_{0,e}=10$ for all streets $e$, leaving all other parameters in equations Eq.~\eqref{outrate} and Eq.~\eqref{eq:decision} unchanged. Vehicles enter the system following a Poisson process with a rate $\nu_\text{in}^0$ with origin and destination sampled from a uniform distribution. When a vehicle enters the system, the driver chooses a path $i$ from one of the possible shortest paths between their origin and destination based on the available (possibly outdated) information, where the travel time along the route $i$ is the sum of the travel times along all streets on this route. As before, route choices depend on the signaled travel times as given by Eq.~\eqref{eq:decision}\,.

\subsubsection{Classifying congested traffic}

Microscopic, agent-based simulations of drivers relying on delayed travel time information for their routing decisions on non-trivial street networks pose several challenges. First, stochastic fluctuations of the number of inflowing vehicles may induce spontaneous and persistent congestion~\cite{krall_number_2020}. Second, as the routes of vehicles typically consist of more than one street segment, a natural additional information delay occurs if drivers make their route choice at the beginning of their trip without later adaptation. Third, classifying a state as congested becomes more subtle in such a stochastic system. Sudden peaks in the street loads can occur \rev{transiently} without inducing permanent congestion. To circumvent wrong characterizations due to random fluctuations in our simulation, we increase the threshold at which we consider the system as congested from $N_\text{max} := N(\nu_\text{max})\approx 16$, which we find from Eq.~\eqref{DGL}, to $N_\text{threshold}=100$ vehicles. We furthermore choose a comparatively long simulation time of 400 time steps to ensure that no eventually congested state is overlooked. 

\subsubsection{Averaging prevents congestion in complex street networks}

The impact of information delay on the collective traffic dynamics on larger street networks is comparable to the impact observed in the minimal two-road setting. The higher the delay, the lower is the critical in-rate for which congestion occurs. This general change in traffic flow dynamics is summarized in the bifurcation diagram \rev{(\fig~\ref{fig:fig2_gridversion}a)}. 

As in the two-road scenario, an increased delay induces higher average loads on the streets in the network. The already existing imbalance between highly frequented central edges and less frequented boundary regions, visible in \fig~\ref{fig:fig2_gridversion}b\,, is reinforced if information is delayed, as demonstrated for $\tau=5$ in \fig~\ref{fig:fig2_gridversion}c\,. 
Beyond a critical delay the street loads increase above $N_\text{max}$, i.e. congestion arises (\fig~\ref{fig:fig2_gridversion}d). Interestingly, the street loads of two alternative routes (marked by red and blue paths with arrows in \fig~\ref{fig:fig2_gridversion}d) undergo oscillations that grow in amplitude, similar to those occurring in the two-road model for high delays (see \fig~\ref{fig:fig2_gridversion}e). Thus, also in non-trivial street networks, the overreaction of drivers induces unbalanced, alternating street loads which eventually lead to a congested state. 

When providing averaged information, the same qualitative changes of the traffic flow dynamics occur for larger grid networks as for the small two-road system \rev{(\fig~\ref{fig:fig3_gridversion})}. \rev{Time-averaging of delayed traffic information stabilizes the free flow state at high delays (a smaller fraction of realizations result in congestion) compared to information on the instantaneous state. For small delays, averaging prevents accurate reaction to current traffic conditions and destabilizes the free flow state (more simulations result in congestion).} 

\subsubsection{Informing only a fraction of drivers prevents congestion}

In a real-world setting, not all drivers rely on (the same) traffic information when choosing their route. To find whether the impact of delay on traffic flow dynamics stays the same for partially informed drivers, we provide only a fraction $f$ of drivers with travel time information, while the remaining fraction $1-f$ decide uniformly randomly for one of the alternative route options. \rev{We then evaluate the critical in-rates at which 50\% of our simulation runs result in congestion.}

If information is almost up-to-date, ($\tau = 1$), \rev{vehicles spread equally across the network as} overreaction is small and hardly impacts the stability of the free flow state (see \fig~\ref{fig:fig2_gridversion}a). On the contrary, the provided information assists drivers to react to current load imbalances and thus prevents congestion. Hence, a reduction of the fraction $f$ of informed drivers as in Fig.~\ref{fig:fig4_gridversion} always induces a decrease of critical in-rates, i.e. an earlier occurrence of instability.

If information is delayed ($\tau \in \{5,15\}$), \rev{drivers that choose routes randomly effectively reduce the overreaction of informed drivers. The route choices of those drivers counterbalance the overreaction to travel time changes. The number of routing decisions that amplify differences in street loads decreases. More precisely,} reducing the fraction $f$ of informed drivers increases the critical in-rate \rev{as long as sufficiently many drivers remain informed}, i.e. the system supports a higher amount of traffic (\fig~\ref{fig:fig4_gridversion}). This effect can be understood in the same way as averaging outdated information. 
\rev{If all drivers collectively chose} the route that drivers are told to be optimal, they \rev{would} cause load imbalance and, eventually, congested states. In a setting where only a fraction of drivers actually have access to the information, this overreaction is damped.  

If \rev{the fraction of informed drivers drops below an optimal value,} too many route choices are made randomly \rev{and} important information on the load distribution within the network is lost\rev{. Intrinsic heterogeneities of street loads, such as the fact that central streets are typically more loaded than outer ones (\fig~\ref{fig:fig2_gridversion}b), result in a more uneven distribution of traffic.} If this general tendency of load imbalance is not known to more than \rev{sufficiently many} the drivers, \rev{congestion arises earlier and the capacity of the system decreases, i.e. congestion emerges} for lower in-rates than if all drivers use the available information. \rev{However, for high delay ($\tau=15$), letting all drivers choose their routes uniformly at random ($f=0$) may still result in a higher capacity than a scenario in which all drivers use the available, out-dated information ($f=1$).}

\section{Discussion}
We have introduced and analyzed a macroscopic model of how delayed travel-time information may affect traffic flow dynamics. Information delays may induce oscillations on alternative routes, both in a simple two-road system and in more complex street networks. Our finding of oscillations due to time delay is in line with previous results modeling traffic flow in a two-street network with agent-based simulations where information is provided with a delay that depends on the street load~\cite{wahle_decision_2000}. Additionally, linear stability analysis of the resulting delay differential equations has revealed that once the delay exceeds a critical value, the load imbalances diverge and congestion arises on both routes. 

We have proposed two possible strategies to prevent overreaction and congestion due to outdated information. First, if travel time information averaged over past observations is provided instead of a value measured at a single time point, the capacity of the system, i.e., the critical in-rate at which congestion occurs, increases for intermediate or high information delays. Thus, this strategy prevents traffic congestion in a similar way in which it prevents destabilizing feedback loops in decentralized smart power grids \cite{schafer_decentral_2015} and reduces blocking probabilities in optical fiber networks \cite{peroza_marval_adaptive_2011}. Second, we found that the collective overreaction is less severe if only a fraction of drivers receives the delayed information. This finding stands in contrast to the na\"ive expectation that providing available information to all individuals may optimally improve system performance. Indeed, our results show that random route choices of the uninformed drivers induce the same increase of the critical in-rate at which congestion occurs as observed for averaged information, similar to previous work that investigated this effect on selfish routing without explicitly discussing the role of information delay~\cite{mahmassaniSystemPerformanceUser1991, liuEffectsInformationHeterogeneity2016, festaModelingImpactOnline2019a}. 

The combination of analytical insights in the minimal two street network and the observations in networks suggests that our results may robustly transfer to more complex settings as well. \rev{While the quantitative details of these results depend on the specific symmetries and parameters of the setting, the mechanism revealed in the minimal example and the grid network is robust for different conditions (see Appendix C).}  
\rev{Similarly, qualitatively the same phenomena may also emerge in other models of traffic flow. For example in cellular automaton models with explicit spatial exclusion, uninformed drivers generally increase the overall travel time for small vehicle densities but may prevent a complete gridlock and increase travel velocities for high densities of vehicles due to random route choice and increased flexibility of drivers \cite{tai_global_2019}.} 
Determining the magnitude of the reported effects in larger empirical street networks is left for future work. However, as \rev{the mechanism is robust and} delays are inherent in providing information in any real world scenario, for example due to updating times and unpredictable feedback mechanisms~\cite{yokoya_dynamics_2004, scellato_traffic_2010}, we expect this effect to be relevant in a broad range of settings.

In summary, our results suggest that providing partially or completely obscured information about travel times may prevent congestion induced by information delays when the delay exceeds a critical value. In the near future, these findings may be especially relevant for autonomous vehicle traffic \cite{chan_advancements_2017}. As traffic flow stability will not be limited by human imperfections such as random braking \cite{nagel_cellular_1992, michael_capacity_1998}, the collective dynamics of vehicle fleets may play a dominant role in the emergence and prevention of congestion.

\begin{acknowledgments}
We thank Dirk Witthaut for valuable discussions. The authors thank the German Academic Scholarship Foundation for its role in initiating the project. V.K. thanks the Evangelisches Studienwerk. M.F.B. thanks the International Max Planck Research School for Intelligent Systems (IMPRS-IS). This research was supported through the Center for Advancing Electronics Dresden (cfaed).\\
\end{acknowledgments}

\section*{Competing interests}
The authors declare no competing interests.

\section*{Data availability}
All material has been made publicly available at GitHub and can be accessed at \url{https://doi.org/10.5281/zenodo.5566330}

\section*{References}
\bibliographystyle{unsrtnat}
\bibliography{references}

\clearpage
\appendix

\section{Linear stability analysis \label{LSA_Appendix}}

\rev{With delayed information, the differential equation governing the street load dynamics is
\begin{align}
    \frac{\mathrm{d} N_i}{\mathrm{d} t}= \dot{N}_i(t) 
    &= F_i(N_i, \left\{ N_j(t-\tau)\right\}_j) \notag\\
    &=\nu_{\text{in},i}\left( 
    \left\{t_{\text{signal},j}\lbrack N_j(t-\tau)\rbrack\right\}_j\right) \notag\\
    &\ \ \ 
    - \frac{N_i(t)}{N_0}\frac{N_i(t)}{t_0\,\exp{\left(N_i(t)/N_0\right)}-1}\ ,\label{complete_model} 
\end{align}
where $j$ indexes all street segments, the rate of inflow $\nu_{\text{in},i}$ into a street $i$ is defined according to Eq.~\eqref{eq:decision},
\begin{align*}
    \nu_{\text{in},i} &= \nu_{\text{in}}^0 \left(\frac{e^{-\beta t_{\text{signal},i}}}{\sum_j e^{-\beta t_{\text{signal},j}}}\right) \ ,
\end{align*}}
and the signalled travel time for road $i$,
\begin{align*}
    t_{\text{signal},i}\lbrack N_i(t-\tau)\rbrack &= t_0\frac{\exp{\left(N_i( t-\tau)/N_0\right)}-1}{N_i(t-\tau)/N_0} \quad,
\end{align*}
depends on the street load at time $t-\tau$\,. \rev{As described in the main text, we set the parameters $t_0 = 1$, $N_0 = 1$ and $\beta = 1$.}

\begin{figure}[ht]
    \centering
    \includegraphics[width=8.5 cm]{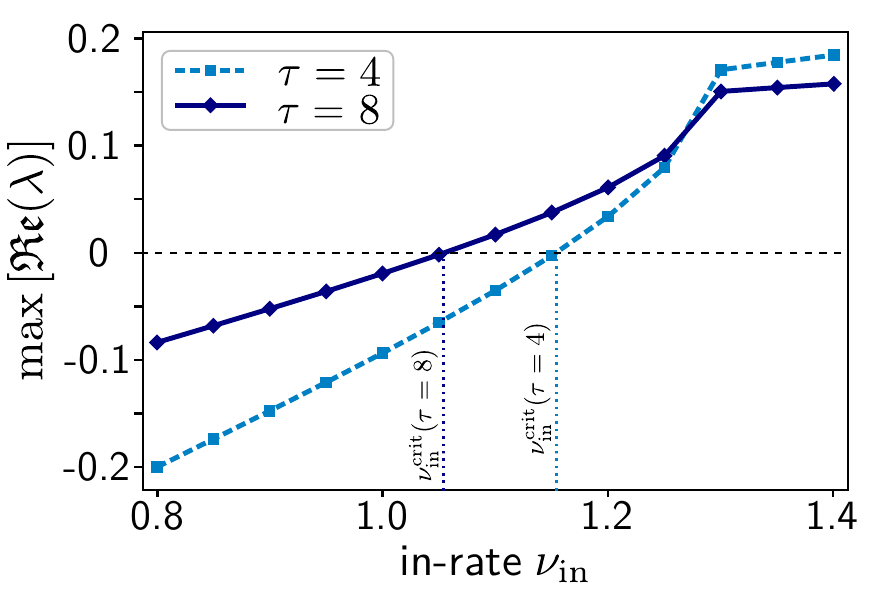}
    \caption{\textbf{Linear stability analysis predicts critical in-rate.} We determine the stability boundary illustrated in Fig.~\ref{fig:fig2_gridversion} and \ref{fig:fig3_gridversion} by finding the critical in-rate $\nu_\text{in}^\mathrm{crit}$ for each information delay $\tau$ above which the traffic flow becomes unstable. This instability is characterized by a solution $\lambda$ with a positive real part of the characteristic equation~\eqref{charEq} (or equivalently Eq.~\eqref{charEqAve} when providing averaged information).
    }
    \label{fig:appendix1}
\end{figure}

To find whether an equilibrium point $N_1^\ast=N_2^\ast=N^\ast$, defined by $\frac{\mathrm{d} N_i}{\mathrm{d} t} = 0$, is stable, we consider a small perturbation from equilibrium\citep{driver_ordinary_1977, roussel_delay-differential_2019}
\begin{align*}
    N_i(t) = N_i^\ast + \delta N_i(t) \quad ,
\end{align*}
linearize the differential equation and insert the exponential ansatz 
\begin{align}
    \delta N_i(t) = A_i e^{\lambda t} \quad. \label{exp_ansatz}
\end{align}
The possible values for the exponent $\lambda$ are determined by the characteristic equation 
\begin{align}
    \mathrm{det}\left(J_0 +e^{-\lambda \tau} J_\tau -\lambda \mathbb{1} \right)
    = 0\quad, \label{charEq}
\end{align}
with \rev{Jacobi matrices $J_0$ and $J_\tau$ with} the entries 
\begin{align*}
    J_{0, ij} &= \left .\left( \frac{\partial}{\partial N_j(t)} F_i(t)  \right) \right\vert_{N_{i,j}(t)=N_{i,j}(t-\tau)=N^\ast}\\
    J_{\tau, ij} &=\left . \left( \frac{\partial}{\partial N_j(t-\tau)} F_i(t) \right) \right\vert_{N_{i,j}(t)=N_{i,j}(t-\tau)=N^\ast}\quad .
\end{align*}
The equilibrium $N^\ast$ is unstable if $\mathfrak{Re}(\lambda) > 0$ for any solution $\lambda$ of equation~\eqref{charEq}\,. Thus, to determine the stability boundary shown in Fig.~\ref{fig:stabchange}, we numerically calculate for each delay the critical in-rate $\nu_\text{in}^\text{crit}$ for which $\mathrm{max}\left[\mathfrak{Re}(\lambda)\right] = 0$ (see Fig.~\ref{fig:appendix1}).

\begin{figure}[h]
    \centering
    \includegraphics[width=8.5 cm]{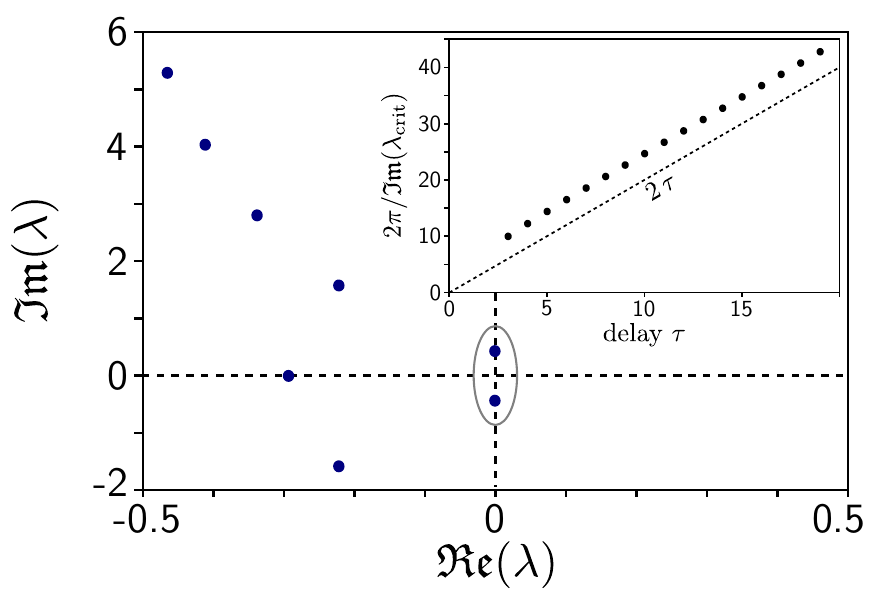}
    \caption{\rev{\textbf{Hopf-bifurcation at the critical in-rate.} (a) Solutions of the characteristic equation [Eq.~\eqref{charEq}] for $\tau=5$ at the critical in-rate $\nu_\text{in}^\text{crit}(\tau=5)\approx 1.115$ in the complex plane. 
    At this bifurcation point, one pair of complex conjugate solutions crosses the imaginary axis (grey ellipse) and the dynamics undergoes a Hopf-bifurcation. 
    (b) Period of the critical oscillations at the bifurcation point as a function of the delay $\tau$, predicted using the imaginary part of $\lambda_c$. 
    } \label{fig:appendix2}}
\end{figure}

\rev{Examining the solutions of the characteristic equation Eq.~\eqref{charEq} in more detail reveals a Hopf-bifurcation where a pair of complex conjugate solutions crosses the imaginary axis (Fig.~\ref{fig:appendix2}): at the critical in-rate, the system transitions from a stable state with decaying oscillations to an unstable state with growing oscillations.
While it is not trivial to a priori predict the exact period of street load oscillations, an analysis of the imaginary part of the critical solutions $\lambda_c$ with maximal real part at the bifurcation point provides valuable insight into the systemic dynamics. 
At the bifurcation point, the pair of critical solutions
\begin{align*}
    \lambda_c = \pm \mathrm{i}\omega
\end{align*}
dominates the dynamics. All other oscillations will decay rapidly since the other solutions of the characteristic equation have negative real part. 
The period of the dominant oscillations at the critical point is thus given by $T = 2\pi/\omega$\,, scaling approximately as $T = 2\tau + \mathrm{const}$. Comparing the results shown in Fig.~\ref{fig:appendix2}b with the exemplary dynamics in Fig.~\ref{fig:stabdelay}d, this approximation seems to indeed provide a useful first estimate of the street load oscillation's period. }

To analyze the stability of the system when providing averaged information, we simplify the integral in equation~\eqref{eq:av_load} by adding a second differential equation for the averaged street loads $N_{i,\text{av}}=\langle N_i(t-\tau)\rangle_{T_\text{av}}$ with two explicit delays $\tau$ and $\tau + T_\text{av}$. The dynamical system is thus described by two delay differential equations for each street $i$,
\begin{align}
    \frac{\mathrm{d} N_i}{\mathrm{d} t} 
    &= G_i(N_i,\left\{N_{j,\text{av}}\right\}_j) \nonumber\\
    &=\nu_{\text{in},i}\left( 
    \left\{t_\text{signal}^j\lbrack N_{j,\text{av}}\rbrack\right\}_j\right) \nonumber\\
    & \ \ -\frac{N_i(t)}{N_0} \frac{N_i(t)}{t_0 \exp{\left(N_i(t)/N_0\right)}-1}\quad, \nonumber\\
    \frac{\mathrm{d} N_{i,\text{av}}}{\mathrm{d} t} =& \frac{1}{T_\text{av}} \left(N_i(t-\tau) - N_i(t-\tau-T_\text{av}) \right)\ . \label{complete_model_averaging}
\end{align}
\rev{with travel time information 
\begin{align*}
    t_\text{signal,i}\lbrack N_{i,\text{av}} \rbrack &= t_0\frac{\exp{\left(N_{i,\text{av}}/N_0\right)}-1}{N_{i,\text{av}}/N_0} \quad,
\end{align*}
based on the delayed, averaged street loads. This model } yields the characteristic equation
\begin{align}
    \mathrm{det}\left(\tilde{J}_0 +e^{-\lambda \tau} \tilde{J}_{\tau_1} +e^{-\lambda(\tau+T_\text{av})}\tilde{J}_{\tau_2}-\lambda \mathbb{1} \right)
    = 0\quad, \label{charEqAve}
\end{align}
where the Jacobians are now 4x4 matrices with entries
\begin{align*}
    \tilde{J}_{0, \alpha \beta} &= \frac{\partial \dot{x}_\alpha (t)}{\partial x_\beta (t)}\\
    \tilde{J}_{\tau_1, \alpha \beta}&= \frac{\partial \dot{x}_\alpha (t)}{\partial x_\beta(t-\tau)}\\
    \tilde{J}_{\tau_2, \alpha \beta} &= \frac{\partial \dot{x}_\alpha(t)}{\partial x_\beta(t-\tau-T_\text{av})}\quad,
\end{align*}
for $x_\alpha, x_\beta \,\in\, \lbrace N_1, N_2, N_{1,\text{av}}, N_{2,\text{av}} \rbrace $\,.

Again, we find the critical in-rate $\nu_\text{in}^\text{crit}$ by determining for each delay $\tau$ (and averaging time window $T_\mathrm{av}$) the \rev{smallest} $\nu_\text{in}$ for which a solution $\lambda$ to Eq.\eqref{charEqAve} with zero real part exists.
\\[12mm]

\rev{\section{Robustness of results for a fraction of informed drivers}}

\rev{The symmetry of the setting plays a subtle role for the results discussed in the main manuscript. For the homogeneous minimal model with two streets of equal length, providing up-to-date information does not improve traffic stability compared to letting drivers decide their routes at random. However, our results on the impact of information in a 5x5 grid suggest that in heterogeneous settings, (not) providing information has a non-trivial effect on overall systemic stability. Here, we illustrate the impact of symmetry on the influence of traffic information that is available to a fraction of drivers with or without delay. }

\begin{figure*}
    \centering
    \includegraphics[width=17 cm]{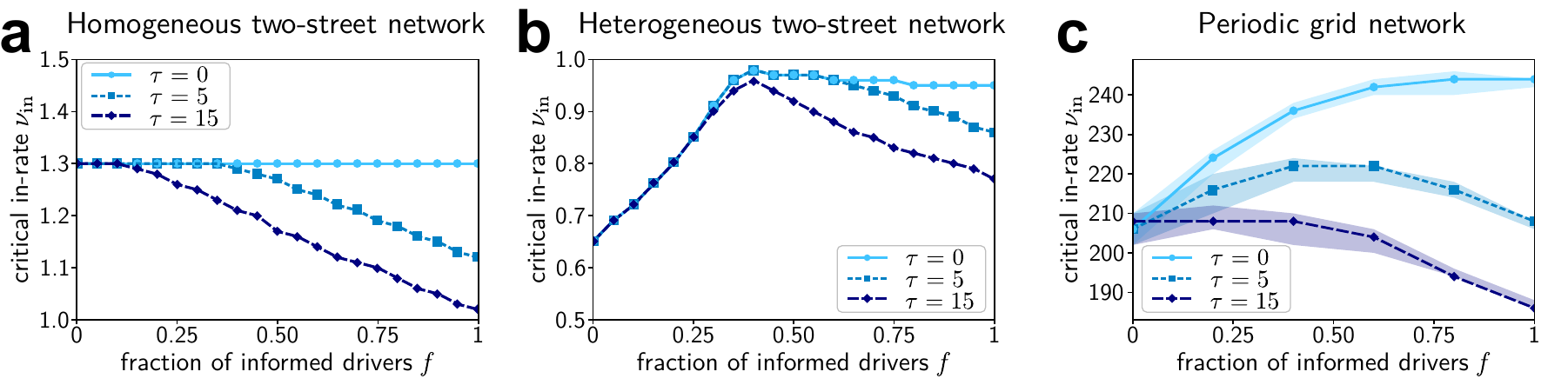}
    \caption{\textbf{Qualitatively similar impact of delayed information across settings.} 
    \rev{(a) In a homogeneous two street network, information provides no benefit as random route choice also results in even street load distribution. Delayed information decreases the critical in-flow as drivers overreact to street load imbalances. Results for direct simulation of the delayed differential equations with initial conditions identical to Fig.~\ref{fig:stabdelay}. 
    (b) In a heterogeneous two street network, uninformed drivers sub-optimally utilize the available street capacity and providing information increases the critical in-rate. If the information is delayed, informing only a fraction of drivers achieves the highest possible in-rates (compare Fig.~\ref{fig:fig4_gridversion}). Results for direct simulation of the delayed differential equations with initial conditions $N_1(t\le0) = N_2(t\le0) = 0.1$.
    (c) Stochastic fluctuations in the microscopic agent-based modeling counter the effect of symmetry. The results for a homogeneous grid with periodic boundary conditions are qualitatively the same as in a heterogeneous setting with fixed boundaries (compare Fig.~\ref{fig:fig4_gridversion}).}
    }
    \label{fig:FIGA3_robustness}
\end{figure*}

\rev{To model a fraction of informed drivers in the differential equation model, we adjust the in-rate such that
\begin{align*}
    \nu_{\text{in}, i} &= \nu_{\text{in}}^0 \left[ f  \left(\frac{e^{-\beta t_{\text{signal},i}}}{\sum_j e^{-\beta t_{\text{signal},j}}}\right) + \frac{(1-f)}{2}\right]\quad ,
\end{align*}
where the informed fraction $f$ of drivers follows the logit model based on the travel time information while the remaining fraction $1-f$ choose each route with equal probability. In the homogeneous setting, random route choice by drivers ($f=0$) equally distributes the street load across both streets. The free-flow state is stable for the highest possible in-flow. Providing information has no additional benefit in terms of the linear stability of the free-flow state, but delayed information may decrease the maximum in-flow where traffic flow remains stable (Fig.~\ref{fig:FIGA3_robustness}a).

However, in a heterogeneous setting with differing free-flow travel times, $t_{0, 1}=1$ and $t_{0,2}=2$, but equal effective capacity $N_{0,1}=N_{0,2}=1$\,, we find qualitatively similar results as for the grid network (Fig.~\ref{fig:FIGA3_robustness}b, compare Fig.~\ref{fig:fig4_gridversion}). 
Even delayed information is partially valuable in a heterogeneous street network, as it enables drivers to avoid the most inferior roads. As long as only an intermediate fraction of drivers is informed, providing outdated information may increase stability.

In more detailed stochastic agent-based simulations as for the 5x5 grid network, street loads in the network inherently fluctuate. This randomness implies that even in grids with periodic boundary conditions, vehicles are not completely homogeneously distributed, in contrast to the deterministic delayed differential equation modeling of the homogeneous two street network. Consequently, we find qualitatively similar results as for the bounded grid in the main manuscript. Reducing the fraction of informed drivers with up-to-date information increases the risk of congestion and informing only a fraction of drivers may be beneficial when information is delayed (Fig.~\ref{fig:FIGA3_robustness}c, compare Fig.~\ref{fig:fig4_gridversion}). 
}

\end{document}